# Photosynthetic Potential of Planets in 3:2 Spin Orbit Resonances

S.P. Brown[1], A.J. Mead[2], D.H. Forgan[1,2], J.A. Raven[3], C.S. Cockell[1]

1 -  UK Centre for Astrobiology, School of Physics and Astronomy, University of Edinburgh, James Clerk Maxwell Building, Mayfield Road, Edinburgh, EH9 3JZ
2 - SUPA, Institute for Astronomy, Royal Observatory Edinburgh, Blackford Hill, EH9 3HJ
3 - Division of Plant Sciences, University of Dundee at TJHI, The James Hutton Institute, Invergowrie, Dundee, UK

Corresponding Author: Duncan Forgan (dhf@roe.ac.uk)

**Abstract**

Photosynthetic life requires sufficient photosynthetically active radiation (PAR) to metabolise. On Earth, plant behaviour, physiology and metabolism are sculpted around the night-day cycle by an endogenous biological circadian clock.

The evolution of life was influenced by the Earth-Sun orbital dynamic, which generates the photo-environment incident on the planetary surface. In this work the unusual photo-environment of an Earth-like planet (ELP) in 3:2 spin orbit resonance is explored. Photo-environments on the ELP are longitudinally differentiated, in addition to differentiations relating to latitude and depth (for aquatic organisms) which are familiar on Earth. The light environment on such a planet could be compatible with Earth's photosynthetic life although the threat of atmospheric freeze-out and prolonged periods of darkness would present significant challenges. We emphasise the relationship between the evolution of life on a planetary body with its orbital dynamics.

# 1 Introduction

The ability to detect extraterrestrial life relies on measurable and well understood biosignatures. Life operating with oxygenic photosynthetic machinery, whereby the host star's energy is the primary energy source and oxygen is the waste product, is a promising biosignature, as well as having the greatest potential for driving primary productivity (Wolstencroft & Raven, 2002 ; Raven & Cockell, 2006), and consequently there has been significant interest into the potential for planets to host photosynthetic life in single star systems (Wolstencroft & Raven, 2002; Cockell et al., 2009).

The habitable zone (HZ) concept, which describes an annulus around a star in which planets of Earth mass and atmospheric composition can sustain surface liquid water (Huang 1959, Dole 1964), has been an extremely useful conceptual tool in understanding under what conditions Earth-like planets may be potentially habitable. However, it is clear that there will be many planets that are not Earth-like in at least one aspect, but still potentially habitable. The most commonly cited example of deviation from Earth-likeness is the rotation rate of the planet. Low mass stars typically possess HZs within the tidal locking radius (cf Dole 1964), and as such it is highly likely that planets of Earth mass inside these zones will eventually become synchronous rotators (i.e. they will enter a 1:1 spin orbit resonance). The climates of planets in synchronous rotation have been studied at length by many authors (Joshi et al., 1997, Joshi 2003, Dobrovolskis 2007, 2009, 2013, Edson et al 2011, 2012, Kite et al. 2011, Yang et al., 2013). However, it is reasonably clear that in general, planets will enter synchronous rotation from a much higher initial spin angular momentum, which is subsequently lost through tidal interactions. During this spin-down period, the planet has a non-zero probability of capture into spin-orbit resonances higher than that of the 1:1 resonance, for example the 3:2 spin orbit resonance.

Within our own Solar System, Mercury is in a 3:2 spin-orbit resonance. The probability of capture into the resonance is increased as a result of Mercury's relatively large orbital eccentricity of 0.206. This high eccentricity also allows the resonance to be sustained through restoring torques on the unchanging (non-tidal) axial asymmetry of the planet (Peale, 1988). These restorative torques keep the longest equatorial axis lined up with the Sun at perihelion.

Mercury itself is unlikely to be habitable, as it lacks an atmosphere and is subject to temperatures in the range 100 - 700°C (Prockter, 2005). However, as this orbital resonance exists within our own Solar System, it could be that such a spin-orbit dynamic is to be found elsewhere. Mercury's capture into this resonance was probable and can be stable over times comparable to evolutionary timescales (Correia & Laskar, 2004, Dobrovolskis, 2007), so it seems possible that M stars may host planets in the habitable zone in 3:2 or other spin-orbit resonances, provided the planet's eccentricity is sufficiently high.

Exoplanets with modest orbital eccentricity appear to be relatively common even at terrestrial planet masses (See Figure 1), and it is conceivable that planets orbiting M stars are captured into a stable 3:2 spin-orbit resonance before or instead of becoming tidally locked. For fast rotating planets, large eccentricity is not necessarily an impediment to potential habitability, depending on the fraction of the planet's orbit that intersects the habitable zone (Kane & Gelino, 2012), and what temperature fluctuations this may generate on the planet's surface (Williams & Pollard, 2002, Dressing et al., 2010).

When the planet rotates sufficiently slowly to be in a spin-orbit resonance, the flux that the planet's surface receives as a function of latitude and longitude becomes non-trivial (Dobrovolskis, 2007, 2009). In this paper, we map the flux patterns generated on the surface of a planet in 3:2 spin-orbit resonance around a low mass star. From these patterns, we speculate on the behaviour of photosynthesising organisms that might inhabit a planet in such a resonance. In section 2 we outline the method for calculating the flux received on the planet's surface; in section 3 we present results for a variety of different orbital eccentricities; in section 4, we discuss the implications of these results for photosynthetic life, and in section 5 we summarise the work.

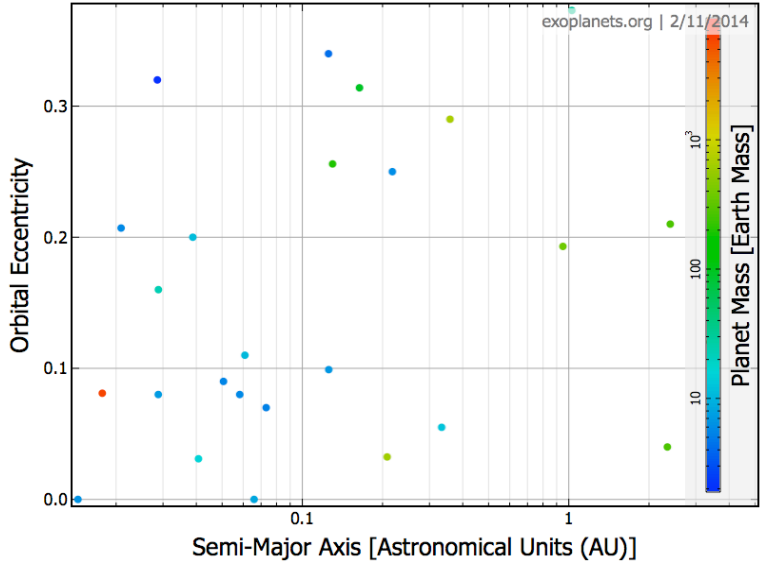

Figure 1: Eccentricity values for known exoplanets orbiting M stars as of February 2014. The planets are colour-coded to reflect their mass in Earth masses (from http://exoplanets.org).

## 2 Method

**2.1 Calculating the Received Flux**

We aim to calculate the flux received at all points on the surface of the planet while orbiting a star in an eccentric 3:2 spin-orbit resonance. Maps of a similar nature were made by Dobrovolskis (2013) - we select a slightly different set of planet parameters to investigate. We specify a fixed Keplerian orbit and spin rotation for the planet, and do not model the various gravitational or tidal forces at play. Over the lifetime of the Sun the tidal reduction of Mercury's average eccentricity (to around 0.2) will incur small changes that are insufficient to disrupt the established spin-orbit resonance (Peale, 1988). Therefore in this first model we can safely ignore tidal spin evolution of the ELP orbit, although we acknowledge that it may play an important role in planetary habitability (Heller et al., 2012).

We assume the planet's orbit is Keplerian, therefore allowing us to characterise it by its eccentricity $e$, and the periastron radius $r_p$. Note how this differs to the usual description of Keplerian orbits in terms of the semimajor axis. In general, Keplerian orbits are elliptical, with the following relation between the radius of the orbit $r$ and the polar angle $\theta$,

$$r = \frac{r_\mathrm{p}(1+e)}{1+e\cos\theta},$$

(1)

and the equations of motion can be integrated to determine the time as a function of $\theta$:

$$\frac{t}{P_\mathrm{o}} = \frac{(1-e^2)^{3/2}}{2\pi} \int_0^\theta \frac{\mathrm{d}\theta'}{(1+e\cos\theta')^2},$$

(2)

in which we set $\theta = 0$ when $t=0$, i.e the initial conditions correspond to the periastron of the first orbit. $P_o$ is the orbital period which is determined via Kepler's Third Law

$$\left(\frac{P_o}{2\pi}\right)^2 = \frac{r_p^3}{GM(1-e)^3}$$

(3)

where $G$ is the Gravitational Constant and $M$ is the mass of the star. We wish to calculate the flux at a given point on a given surface element of the planet denoted by its latitude and longitude, $\nu$ and $\beta$ respectively. We define a fixed unit vector normal to that surface element, **p**, which will rotate along with the planet's spin:

$$\hat{\mathbf{p}} = \cos\nu\cos\beta\,\hat{\mathbf{x}} + \cos\nu\sin\beta\,\hat{\mathbf{y}} + \sin\nu\,\hat{\mathbf{z}},$$

(4)

This vector will determine what is "seen" by a given point on the surface of the planet.
We define the equator as $\nu=0$. As the planet spins, the apparent value of $\beta$ relative to a stationary coordinate system changes as

$$\beta = \beta_0 + \frac{2\pi t}{P_s},$$

(5)

Where $P_s$ is the rotation period of the planet, and $\beta_0$ is the longitude at $t=0$, i.e. the apparent longitude measured at periastron (see Figure 2). We assume the planet has no axial tilt, and hence $\nu$ remains a constant.

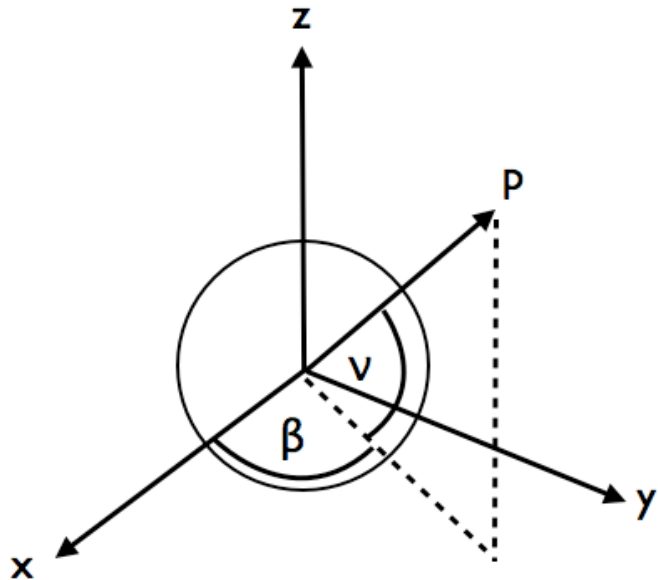

*Figure 2: Illustrating longitude and latitude coordinates $\beta$ and $\nu$ respectively.*

As we assume no axial tilt, we may write the unit radial vector which describes the orbit as

$$\hat{\mathbf{r}} = \cos\theta\,\hat{\mathbf{x}} + \sin\theta\,\hat{\mathbf{y}},$$

(6)

and taking the dot product of this with the unit vector on the surface of the planet (equation 4) gives the cosine of the angle $\pi$-$\varphi$ between the point on the surface of the planet and the star at a given time:

$$\cos\phi = -\cos\nu\cos\beta\cos\theta - \cos\nu\sin\beta\sin\theta = -\cos\nu\cos(\beta-\theta)$$

(7)

The radiation flux $F$ received at a point on the surface of the planet is then given by

$$F = \frac{L\cos\phi}{4\pi r^2},$$

(8)

where $L$ is the intrinsic luminosity of the star. In general this will vary over the course of an orbit both because the distance between the planet to the star changes due to the orbital eccentricity, and because the observed position of the sun in the sky, described by $\phi$, would also change in a non-trivial way as the planet orbits the star. Note that where the cosine of $\phi$ is less than zero, the planet's surface is in darkness, and therefore the flux is set to zero.

**2.2 The Flux Available for Photosynthetic Organisms**

Oxygenic photosynthesisers on Earth make use of a range of wavelengths which are constrained by a number of evolutionary, environmental and cell energetic factors (Wolstencroft & Raven, 2002; Falkowski & Raven, 2007; Kiang et al., 2007a, b; Stomp et al., 2007; Raven, 2009a; Bjorn et al., 2009; Milo, 2009). This range of wavelengths is centred on a wavelength value a little below that of the peak wavelength (Wolstencroft & Raven, 2002; Raven, 2011). We will assume that photosynthesisers relying on radiation from M stars will adapt to utilise flux around the peak wavelength also, where the photosynthesis mechanism may rely on extra photons to compensate for the deficiency in individual photon energy at such long wavelengths (Hill and Bendall 1960, Hill and Rich, 1983, Heath et al 1999).

We can estimate the flux density of photons suitable for photosynthesis by assuming that the star emits all its energy at the peak wavelength, giving

$$F_{\mathrm{PFD}} = \frac{L\cos\phi}{4\pi r^2}\frac{\lambda_{\max}}{hc}$$

(9)

where $h$ is Planck's constant, $c$ is the speed of light and $N_A$ is Avogadro's constant (McDonald, 2003; Puxley et al., 2008). The peak wavelength can be determined via Wien's Law, giving

$$F_{\mathrm{PFD}} = \frac{L\cos\phi}{4\pi r^2}\frac{b}{hcT}$$

(10)

where Wien's constant $b = 2.8983 \times 10^{-3}$ *mK*, and $T$ is the star's effective temperature. The peak wavelength for our star is $\lambda_{max} = 783$ *nm*. Alternatively, we could integrate over the entire blackbody curve instead of using only the peak, giving the bolometric photon flux density to be

$$F_{\mathrm{PFD}} = \frac{L\cos\phi}{4\pi r^2}\frac{30\,\zeta(3)}{\pi^4 k_B T}$$

(11)

where $\zeta$ is the Riemann Zeta function, and $\zeta(3)$ is approximately 1.202. The ratio of equations (10) and (11) is a constant, approximately 0.543, i.e. the integrated photon flux density is around twice that of the flux-at-peak approximation. Equation (10) is used throughout.

Oxygenic photosynthesis requires a suitable atmosphere and sufficient light in wavelength ranges that are photosynthetically active. Plants require a minimum level of atmospheric $CO_2$ concentration of approximately 10 p.p.m (Caldeira & Kasting, 1992), unless a more effective method of carbon acquisition for photosynthesis is used (Bar-Even et al., 2010, Bar-Even et al.,2012). A suitable atmosphere must meet certain requirements including the presence of $O_2$, $N_2$ and a minimum partial pressure of $CO_2$, to control planetary heating by the greenhouse effect (Lovelock & Wild, 1982; Caldeira & Kasting, 1992). Assuming a sufficient atmosphere, Photosynthetically Active Radiation (PAR) will be modified through attenuation by the atmosphere and also by water, for aquatic organisms. However, we cannot comment on these effects, as we do not attempt a radiative transfer calculation in this work.

**2.3 Effects of the 3:2 spin orbit resonance**

So far our analysis has been general (at least, for a planet without axial tilt) but now we restrict ourselves to consider a planet in a 3:2 spin-orbit resonance, i.e.

$$3P_s = 2P_o .$$

(12)

This system is completely symmetric every two orbits ($2P_o$) so that it is sufficient to consider the evolution of just two orbits. A point facing in the direction of the star at periastron at the beginning of the first orbit will be facing directly away from the star at the beginning of the second orbit and then directly towards the star again at the beginning of the third orbit.

The star will be observed to undergo retrograde motion on the sky as viewed from the planet if the planet's orbit is faster than the planet is spinning in terms of angle change per time, or equivalently

$$\dot{\theta} > \dot{\beta}$$

(13)

In other words, if the above equation is satisfied, the star's motion across the sky will change direction. If the star is setting in the sky during this transition, the sunset may turn into a sunrise, and the 'day' will be significantly lengthened. Equally, a rising sun may immediately set again, cutting the day short.

In our analysis, we fix the spin rate as a constant, but the rate of change of orbit angle depends on the angle itself (as well as the eccentricity). We will see that above a critical eccentricity, the above condition for retrograde motion can be satisfied.

The spin rate of the planet is simply

$$\dot{\beta} = \frac{2\pi}{P_s}$$

(14)

The rate of change of orbit angle $\theta$ is given by

$$\dot{\theta} = \frac{2\pi}{P_o} \frac{(1+e\cos\theta)^2}{(1-e^2)^{3/2}},$$

(15)

This function clearly has a maximum at periastron ($\theta = 0$) so to find the critical value of $e$ which allows retrograde motion, we equate equations (14) and (15), and set $\theta = 0$:

$$\frac{2\pi}{P_s} = \frac{2\pi}{P_o} \frac{(1+e)^2}{(1-e^2)^{3/2}},$$

(16)

which, given that $3P_s = 2P_o$, can be solved to give $e \approx 0.191$. Hence, any eccentricity larger than this critical value will result in retrograde motion of the star on the sky. Figure 3 shows this effect. The panels shows the height of the star in the sky as a function of time for several latitudes. In the left panel, the eccentricity is zero, and the star moves continuously from one horizon to the other (-90 degrees to +90 degrees respectively). In the left panel, the eccentricity is 0.3, and exceeds the critical eccentricity for retrograde motion. The star changes direction in its motion across the sky. At longitudes of zero degrees, this is sufficient for the sun to set, rise and set again at an orbital phase of ~1.

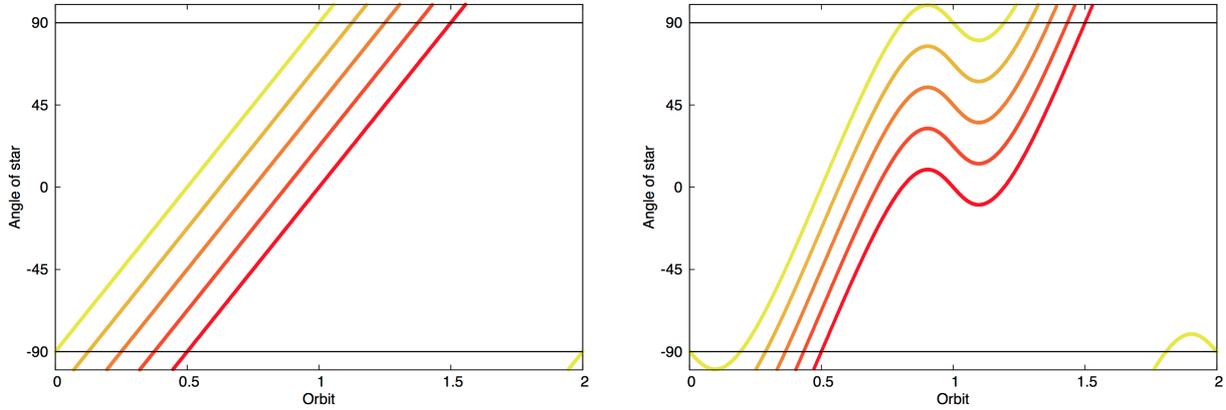

*Figure 3: The position of the star in the sky over 2 orbits, at longitudes along the equator of 90 degrees (yellow), 67.5, 45, 22.5, 0 degrees (red), as a function of orbital phase (x-axis). The y-axis shows the star's zenith angle - an angle of 0 represents the star at its highest in the sky, and angles of +/- 90 degrees represents the horizons. The left graph represents an orbit of eccentricity e=0, and the right represents an eccentricity of e=0.3.*

The tidal forces that drive a system into a spin-orbit resonance will typically drive the system into a Cassini state, which can reduce the obliquity of the planet to be zero (Biswas, 2000) and this is the case for Mercury (Margot et al., 2012). We therefore consider only the case where the axial tilt is zero, but we note that high obliquity states are also possible (see Discussion).

### 2.4 Initial Conditions

In our analysis we consider an ELP which we assume to have a similar level of water cover and atmosphere to the Earth, and consequently a similar albedo and greenhouse effect. However, we assume that the obliquity is zero, unlike the Earth's average tilt of approximately 23.5 degrees. We select an M star of mass 0.3 solar masses, effective temperature 3500 K, and a luminosity of 0.01 times the solar luminosity. We fix the periastron radius $r_p$ as 0.1 AU

(where 1 AU is the distance from the Earth to the Sun). The habitable zone for a star with these parameters extends from 0.1 to 0.2 AU (Kopparapu et al 2013), placing the planet close to the inner edge at periastron. Note that increasing *e* will increase the semi-major axis, and ergo the period of the planet's orbit. Also, increasing e will decrease the fraction of the orbit spent in the habitable zone, as displayed in Table 1. However, we should be careful to note that habitable zone boundaries are calculated assuming the planet's rotation period is 24 hours, which is much shorter than modelled here, and it is unclear to what extent such boundaries can be considered valid in this case.

We explore both latitudinal and longitudinal variations in insolation. The flux on non-zero latitudes (i.e. latitudes above or below the equator) can be simply found by multiplying by a factor of cos *v* (see equation 9).

| eccentricity | Orbital Period (Days) | Fraction of Orbit in HZ |
|---|---|---|
| 0 | 21.1 | 1 |
| 0.2 | 29.5 | 1 |
| 0.4 | 45.3 | 0.56 |
| 0.5 | 59.6 | 0.34 |
| 0.7 | 128.0 | 0.12 |
| 0.8 | 236.0 | 0.06 |

*Table 1: The orbital period of the ELP as a function of its eccentricity, given the periastron radius is fixed at $r_p$ = 0.1 AU. The right hand column notes what fraction of the orbit is spent in the habitable zone, which extends from 0.1 to 0.2 AU (Kopparapu et al 2013).*

# 3 Results

Figure 4 shows maps of the total received flux as a function of longitude and latitude, over 2 orbits of the planet, for various eccentricities. If the orbit is circular (top left plot), the orbital velocity remains constant, and the flux is distributed symmetrically across the surface. However, the 3:2 spin orbit resonance seems to require a modest eccentricity to be sustained. Increasing the eccentricity to 0.2 (top right) is sufficient to produce retrograde motion of the star in the planet's sky, with the received flux being concentrated around closest and furthest approach from the star, producing two flux hotspots on the planet's surface, centred at longitudes equal to 0 and 180 degrees (as these correspond to the longitudes facing the planet at closest during orbits 1 and 2).

These hotspots become more focused, with a reduced footprint in longitude as the eccentricity is increased to 0.4 (middle left plot). However, once the eccentricity is increased to 0.5 (middle right), the flux received at longitudes of 90 and 270 degrees begins to increase, until at around e=0.7 (bottom left), the integrated flux distribution looks remarkably similar to that of the e=0 case. However, we should note that increasing the eccentricity increases the time the planet spends at apastron, and consequently reduces the total integrated flux received by the planet's surface (Dressing et al 2010). Increasing the eccentricity to e=0.8 (bottom right) shifts the phase of the hotspots to 90 and 270 degrees, which now correspond to the points where the planet faces the star in between periastron and apastron.

We can see from these maps that the time-averaged longitudinal distribution of light on the planet surface is intimately connected to both the orbital and spin phase of the planet. But how does this affect the "days" and "nights" experienced by organisms at different longitudes on the planet surface?

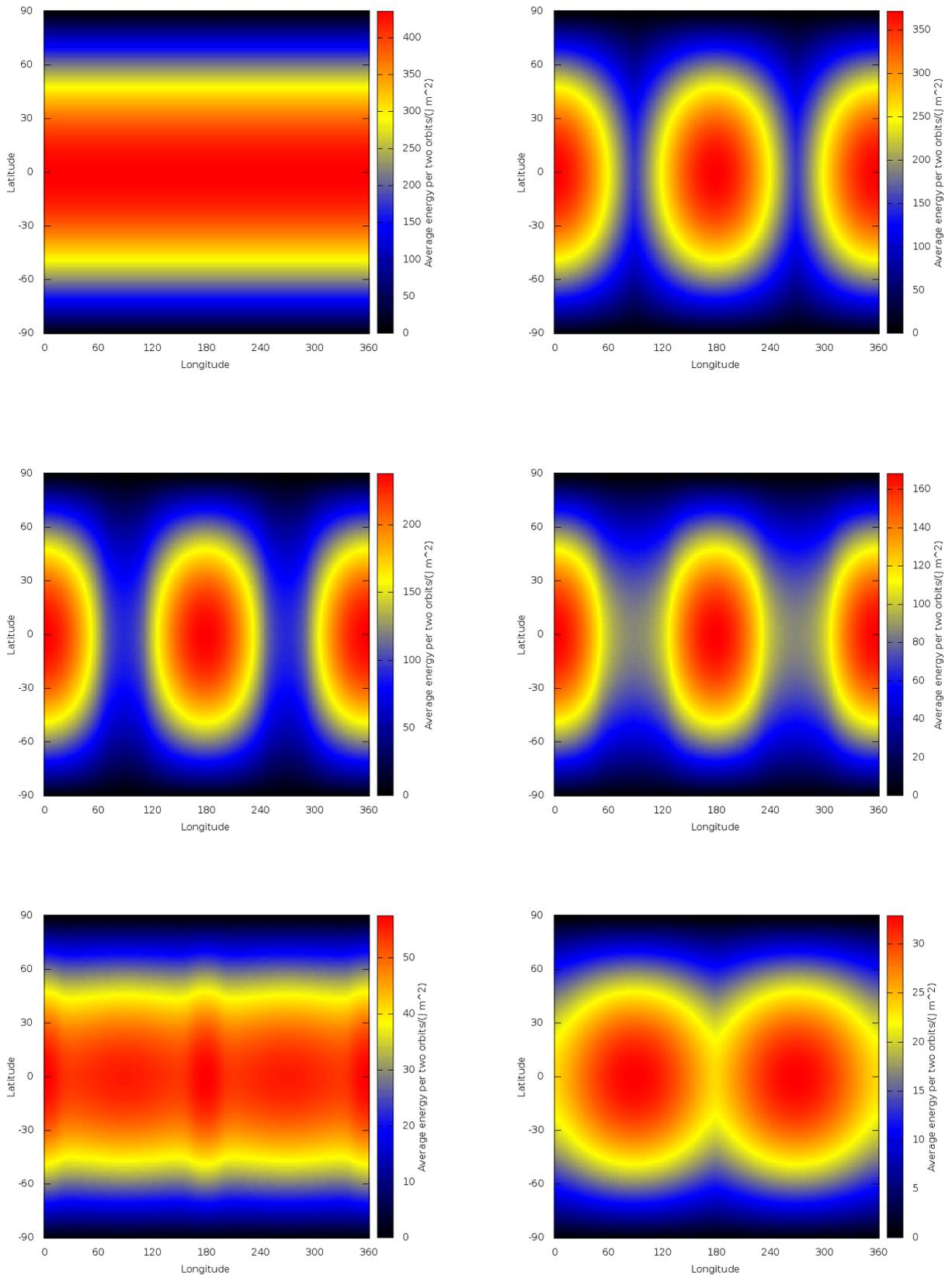

*Figure 4: Integrated energy received over 2 orbits as a function of longitude (x-axis) and latitude (y-axis) for various orbital eccentricities. The graphs are normalised so that 1 represents the maximum energy, and 0 the minimum. The graphs represent from the top, moving left to right to the bottom, the eccentricities e=0, e=0.2, e=0.4, e=0.5, e=0.7, e=0.8. At periastron, the longitudes facing the star will be either 0 or 180 degrees depending on orbital phase.*

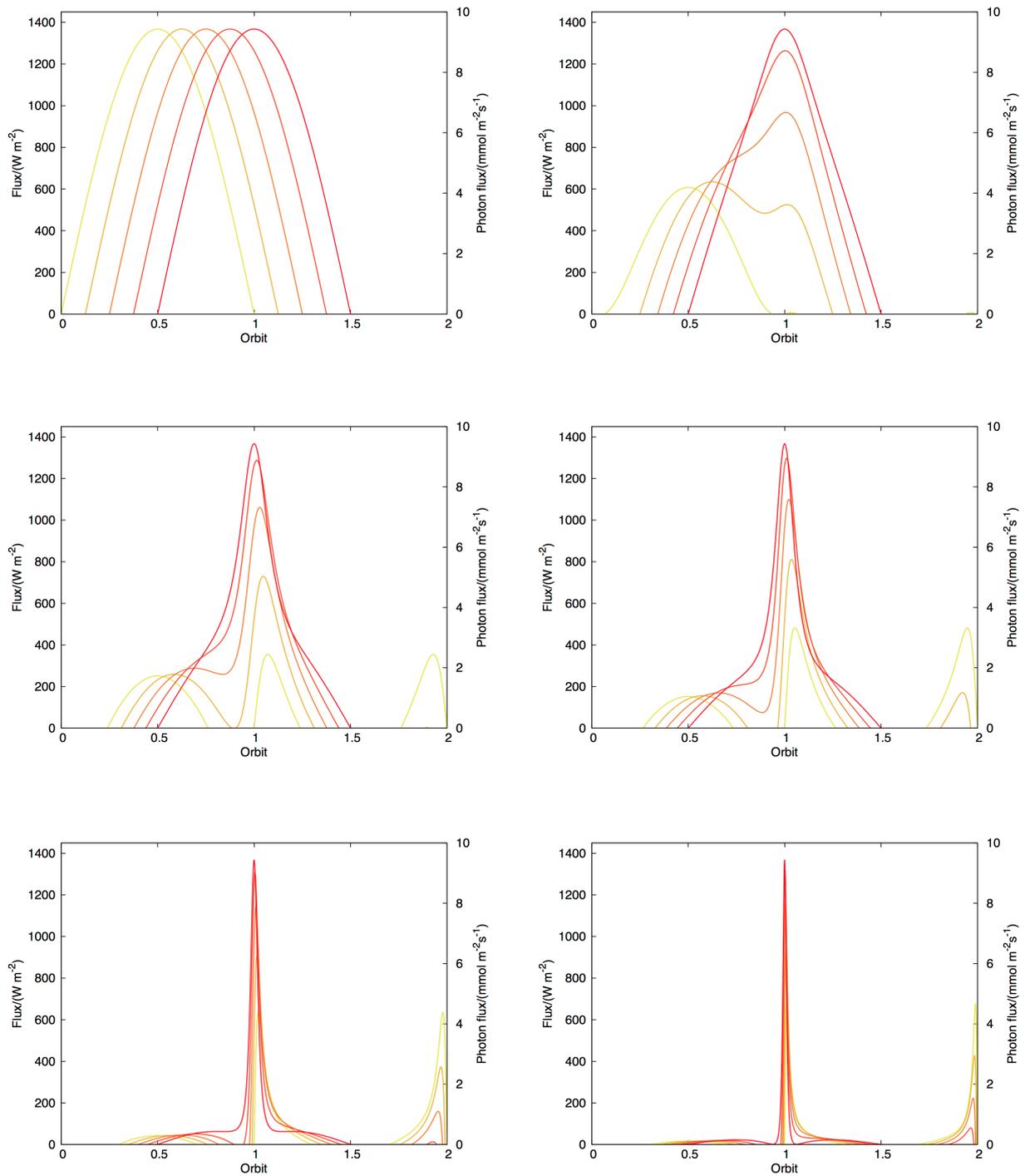

*Figure 5: Flux received over 2 orbits at longitudes along the equator of 90 degrees (yellow), 67.5, 45, 22.5 and 0 degrees (red), as a function of orbital phase (x-axis). The left y-axis shows the flux in W m$^{-2}$, and the right y-axis shows the photon flux density (PFD) in mmol m$^{-2}$ s$^{-1}$. The graphs represent from the top, moving left to right to the bottom, the eccentricities e=0, e=0.2, e=0.4, e=0.5, e=0.7, e=0.8. The curves for longitudes greater than 90 degrees are either mirrored around orbital phase=1 (90-180 degrees) or out of phase with the above (180-360 degrees). Note that the planet passes periastron at orbital phases of 0, 1 and 2.*

Figure 5 shows how the received flux varies at different longitudes on the planet's equator as a function of orbital phase. We do not plot curves for longitudes greater than 180 degrees, as the flux patterns are either mirror images or the same pattern, but out of phase with those plotted. In the circular orbit case (top left plot), the planet's daytime flux is uniform across all longitudes, as the planet's orbital velocity is constant and its obliquity is zero. Increasing the eccentricity to 0.2 (top right plot) introduces retrograde motion, producing local minima in the flux curves at 67.5 degrees longitude. This is evidence of the star changing direction on the sky, setting and rising on the same horizon. As the eccentricity is increased to 0.4 and 0.5 (middle plots) this effect becomes pronounced and extended to higher longitudes, effectively adding an extra "day" for observers close to 90 and 270 degrees longitude, with peak fluxes much lower than the daytime flux at 0 degrees.

As the eccentricity increases above 0.7, the flux received at 0 degrees longitude becomes highly peaked around periastron passage, and as such the integrated flux that this longitude receives must decrease. The flux received at 90 degrees longitude is in general lower in magnitude, but is distributed over a greater timespan (as it now has an extra "day" towards the end of the second orbit). This appears to be why the maps in Figure 4 show the "hotspots" shifting from 0 and 180 degree longitudes to 90 and 270 degree longitudes. Note that the maximum flux received at any instant as described in Figure 5 does not change with eccentricity. This is an important distinction - increasing eccentricity reduces the time-averaged flux a latitude may receive, but not the maximum instantaneous flux.

# 4 Discussion

We have calculated different light cycles on the 3:2 resonant ELP than experienced on Earth. These light cycles, which most notably feature longitudinal differentiations as a function of eccentricity, present a set of novel environmental challenges.

With this variability of light environment in mind we note that Williams & Pollard (2002) have argued that habitability depends primarily on the average stellar flux incident over a full orbit. Each of the ELP-star scenarios receive an acceptably high average PFD for phototrophs, but the fluctuations around this average can be quite large depending on the orbital eccentricity and longitude. High latitude planetary locations received low levels of PAR, although not unacceptably low for marine photolithotrophs (> 10 $n\,mol\,m^{-2}\,s^{-1}$, Raven et al., 2000).

The environmental challenges for phototrophs, associated with slow planetary rotation, and the longitudinal differentiation of this calculated radiation environment are considered.

**4.1 Atmospheric Freezeout**

When a night lasts so long and, consequently, the period of rotation is slow, it is reasonable to make some approximation to tidally locked systems. Although a tidally locked system has a perpetual 'dark-side', the speed of the planet's rotation is much less than the circulation of a tidally locked atmosphere, according to models (Merlis & Schneider, 2010), allowing for some form of approximation. Of course, the atmospheric circulation will be determined by the individual characteristics of the planet and by insolation. These tidally locked models typically assume very low or zero-eccentricity orbits. Therefore, when any comparisons are made between the 3:2 spin orbit resonance and tidally locked systems this disparity should be kept in mind.

On a tidally locked planet there is the risk of an irreversible condensation of the atmosphere on an ice sheet covering the perpetually dark side. It has been shown, through detailed modelling, that if the atmosphere is sufficiently dense and opaque in the IR, heat transport to the dark side will be sufficient to avoid such a collapse of the atmosphere (Joshi et al., 1997; Joshi, 2003). When a habitable planet is near the edges of the CHZ then this condition is fulfilled as the atmosphere will contain high partial pressures of either $H_2O$ or $CO_2$ (Selsis et al., 2007). The resultant conditions on such a planet would not be Earth-like, as the atmosphere would be super-rotating (Selsis et al., 2007), although not an overwhelming obstacle to planetary habitability. Super rotation would certainly be advantageous for heat transfer and a unidirectional atmospheric transfer of organisms. The effective planetary temperatures calculated do not consider the efficiency of heat transport around the planet. It is likely that there would be notable temperature differences between the illuminated and dark side of the ELP. If planets trespass outside of the CHZ in an orbit, even for a fractional period of their orbit, then the chances of irreversible condensation on the dark-side are increased (Selsis et al., 2007).

It has been suggested (Selsis et al., 2007) that a tidally locked planet could remain habitable, without a permanent freeze-out on the dark side. This relies on the facilitation of greenhouse warming by gases remaining gaseous at low temperatures, for example methane. Kite et al. (2011) provide a discussion of how small changes in the planetary atmosphere's pressure can induce climate instability on tidally locked exoplanets; illustrating the sensitivity and complexity of such a system. Investigations into weathering and the equilibrium atmospheric $CO_2$ level have indicated that the positioning of the substellar point relative to the continents plays an important role in tidally locked planet habitability. When Edson et al. (2012) altered the location of the substellar point from over the Atlantic Ocean to the Pacific Ocean on the Earth they observed that a further quarter of the planet's surface had become habitable. From the conclusions of Edson et al. (2012) it would appear that habitability is not greatly restricted by tidal locking. Therefore, it would be expected that a 3:2 orbital resonance would pose even less of a constraint on habitability.

## 4.2 Prolonged Darkness

Many terrestrial photosynthetic organisms can store enough energy to last through 180 Earth days of darkness, even in warm conditions where metabolic energy consumption is quite rapid (Beerling & Osborne, 2002; Royer et al., 2003; Brentall et al., 2005).

On Earth, oxygenic photosynthesisers are exposed to stellar radiation far more frequently than those on the considered ELP would be. Despite this, some phytoplankton are known to spend a long period of time resting in sediment, more so than time spent growing in the water column (Fryxell, 1983). Some phototrophs could endure extended periods of darkness by displaying similar tendencies for dormancy, as either a morphological or a functional trait.

This argument is restricted to parts of the continental oceanic shelf and shallow freshwaters. In the ocean, and in the 25 million year old Lake Baikal, at its maximum depth of 1636m (Ryves et al., 2003; Jewson et al., 2008), only coastal phytoplankton have specific resting stages. The open (deep) water species do not have such resting stages (Ryves et al., 2003; Jewson et al., 2008). The Lake Baikal data illustrates that the phenomenon has evolved in fresh waters, functionally if not morphologically, and is not restricted to marine habitats. Re-suspension from sediment relies on upwelling from the surface of the sediment up to waters with sufficient illumination for photosynthesis. These mixing depths are only a few tens of metres in permanently stratified tropical waters, but seasonally can be as much as 800m in the North Atlantic at 61 degrees N in February-April (Falkowski & Raven, 2007). Given that the average ocean depth is roughly 3.5km, it is evident that only a small minority of the ocean is suitable for the necessary re-suspension upwelling. Oxygenic photosynthesis evolved in the Earths' waters over 2.4 billion years ago (Blank & Sanchez-Baracaldo, 2010), surviving major extinction events over this period.

This indicates that some coastal phytoplankton groups (Ribeiro et al., 2011) have the capacity to survive, to some extent, adverse planetary habitability conditions, with the ability to repopulate when conditions become favourable once more. Catastrophic events on Earth, such as asteroid impacts, have plunged the planet into darkness followed by an extended period of greatly reduced stellar light reaching the surface (Ribeiro et al., 2011). During the hypothesised "Snowball Earth" it is supposed that several refuges would have been available to the pre-existing eukaryotic algae, protists and testate amoebae who survived (Moczydlowska, 2008). Among these suggested refuges are tectonic hot spots, responsible for a small amount of upwelling, induced by hydrothermal vents.

Photosynthetic organisms with resting stages are fit to tolerate extended periods of darkness. The resting times of several phototrophs have been investigated *ex situ* (phytoplankton not in marine sediment samples) and times of months to years have been reported (Hargreaves et al., 1983 ; Lewis et al., 1999), while in natural marine sediments times of the order decades have been recorded (Keafer et al., 1992; McQuoid et al., 2002; Mizushima, 2004). Work on diatoms and dinoflagellates (Lewis et al., 1999) stored in containers of sediment at 5°C revealed that many species survived at least 27 months. The growth performance of germinated cells after dormancy is thought to be reduced due to energy losses incurred by this state of dormancy (Ribeiro et al., 2011). However Ribeiro et al. (2011) reported phytoplankton viability after 87 ± 12 years in low oxygen silt fjord sediment. They cultured *Pentaphastsodinium dalei*, a dinoflagellate, observing a growth performance unaffected by their dormancy of approximately 87 years.

There are several known algae with genetically integrated photolithotrophy as their common mode of nutrition which can also express an ancestral nutritional mode of phagotrophy with digestion intracellularly (Jones, 1994; Jones et al., 2009; Raven et al., 1997, 2009b, Flynn et al., 2012). See Raven et al. (2009b) for a fuller account of the species with this property, which include green algae (*Pyramimonas*). Such algae are potentially mixotrophic, the extent to which both modes of nutrition are utilised depends on several factors; the capacity for each trophic mode in each alga, the environmental conditions and on whether the alga require an obligatory level of photosynthesis to persist (Raven et al., 2009b). It was found, through the modelling of planktonic mixotrophs, that an obligatory amount of photosynthesis was required for the organism to persist (Flynn & Mitra, 2009). Despite that conclusion, laboratory experiments (Jones et al., 2009) demonstrated that mixotrophic protists could survive 6 months of darkness, quickly resuming photosynthesis when illuminated once more. There is some evidence suggesting that a mixotrophic organism could potentially survive greatly extended, by Earth standards, nights and days. Of course, other suitable environmental conditions such as an availability of inorganic nutrients, as required for photosynthesis, and suitable organic matter, to be consumed phagotrophically, must be upheld.

Phototrophs have shown resilience in laboratory experiments and *in situ*; presently and over the course of the Earths' biotic history. Such biota are of particular relevance to these investigations if their survival in the dark can be prolonged by mixotrophic and/or dormancy behaviours. As the periods of darkness on the modelled ELP are of order of months rather than years, these behaviours seem like promising coping mechanisms for life on a spin-orbit resonant planet

## 4.3 Longitudinal Position on the planet

As we have already seen, the ELP's light environment depends sensitively on longitudinal position and eccentricity. If we consider the case where e=0.4 (Figures 4 and 5, middle left plots), we can see clearly that to maximise the PFD received, there are preferred longitudes of 0 and 180 degrees. If the planet preserved a non-zero obliquity, then there is the potential for preferred latitudes beyond that of the equator (Dobrovolskis 2009).

Longitudinal positioning could therefore play a critical role in planetary habitability calculations. It has been suggested that plants segregate according to an array of environmental niche axes, including gradients of light (Silvertown, 2004) and may do so on the ELP with respect to longitude as well as latitude and depth.

If we assume a sustained periodic light environment over evolutionary timescales then it can be supposed that, if similar life were to evolve, endogenous biological clocks corresponding to the periodicity of that light-dark cycle would evolve. Related differences to plant behaviour, physiology and metabolism asssociated with the indpendently evolved biological clock would be expected.

On Earth, it is accepted that photoperiodism is a major regulator in plant behaviour with respect to latitudinal position (Thomas & Vince-Prue, 1997). In this case, we suggest an extension of this photoperiodism as a regulator of plant behaviour as a function of longitude. Known stationary photosynthetic organisms would presumably adapt to these irradiation patterns, as they have adapted to them on Earth, to maximise their capacity to photosynthesise.

**4.4 The impact of slow rotation on radiation environment**

A planet in 3:2 spin orbit resonance rotates slowly, resulting in a reduced magnetic moment of the global dipole moment. Considering the case of the ELP orbiting an M star, there may be drastic implications for the habitability of the planet. At perihelion the planet will be exposed to the maximum stellar flux, including solar flare cosmic rays (CR). Equally, orbiting closer to the host star is advantageous in avoiding exposure to galactic CR which are largely blocked by the astrosphere (Dartnell, 2011). Ionizing radiation from CRs is responsible for planetary evaporation (Lammer, 2003; Poppenhager, 2012) and can be severely detrimental to photosynthetic life, including effects such as DNA damage (Britt et al., 1996).

M stars have a high X ray and UV activity (high energy), up to ages of a few Gyr (Selsis et al., 2007; Lammer, 2009). Planetary candidates around M stars can be exposed to XUV fluxes 1 to 2 orders of magnitude higher than their counterparts around solar type stars (Selsis et al., 2007).

Plants can repair and protect themselves against UV radiation, to an extent, by means including protective pigments, which can be on or around algae, and chloroplast movement (Barnes et al.,1987; Park et al., 1996). If photosynthetic organisms live in a marine environment, or within substrates such as rock, then they can satisfactorily avoid strong UV flares (Cockell et al., 1999; Cockell et al., 2009). Planets with thick cloud cover may offer extra UV protection (Mayer et al., 1998). It is thought that a dense atmosphere should be protective enough to provide protection from enhanced solar flare CR, however Dartnell (2009) stresses that the consequences of extremely energetic particle events should not be neglected as they could pose as a direct radiation hazard at the planetary surface.

Thus it may be that photosynthetic organisms could be restricted to a cross section of latitudes, longitudes and depths which satisfy PAR requirements and are suitably shielded from high energy radiation.

**4.5 Limitations of the Model**

**4.5.1 The Effects of General Relativity**

This model assumes the ELP orbits with a fixed set of Keplerian parameters. However, we know that Mercury, our Solar System's example of 3:2 spin-orbit resonance, is subject to precession of perihelion as a consequence of General Relativity (GR). It is therefore reasonable to assume that GR effects (not yet considered) may play an important role in spin-orbit resonant ELPs.

In standard GR, orbits in the Schwarzschild metric are mediated by a force with the form

$$\ddot{\mathbf{r}} = -\frac{GM}{r^2}\hat{\mathbf{r}} - \frac{3GMh^2}{c^2 r^4}\hat{\mathbf{r}}$$

(17)

where h is the specific orbital angular momentum. Provided that either r is small, or h is large, the perturbative effect of the second term becomes significant. Averaging this perturbation over an orbit shows that the magnitude of the eccentricity is unaffected, but the orbit will precess, with a period

$$P_{\rm p} = \frac{2\pi}{3} \sqrt{\frac{r_{\rm p}^5 c^4 (1+e)^2}{G^3 M^3 (1-e)^3}}$$

(18)

or in a more palatable form

$$P_{\rm p} \approx 31.5 \sqrt{\frac{(r_{\rm p}/{\rm AU})^5 (1+e)^2}{(M/M_\odot)^3 (1-e)^3}} \text{ million years}$$

(19)

The ELP investigated here would therefore experience precession of perihelion on timescales between 5 to 200 million years depending on the eccentricity. As the perihelion precesses, the hotspots on the planet's surface will precess on the same timescale, shifting the preferred location of biomes as it does so. Depending on the landmass and ocean distributions of the ELP, organisms retreating towards the coast may suddenly find their niche disappearing, with nowhere to escape to.

However, given that these timescales are of the same order as evolutionary timescales, and they are also somewhat larger than those of the Milankovitch cycles that Earth experiences, which are typically of the order of 10 to 100 thousand years (see e.g. Berger et al 2005, Spiegel et al 2010), then it seems reasonable that the rate of precession of perihelion does not rule out habitability, although it will provide a strong selection pressure for future speciation, and in extreme cases perhaps cause extinctions.

**4.5.2 Orbital Perturbations from other Planets**

Another source of perturbation to the ELP's orbital parameters comes from neighbouring planetary bodies. It is unclear to what extent these perturbations change the probability of capture into a resonance. This being said, it is likely that even with successful capture into resonance, the presence of other planets may introduce eccentricity variations, and precession of periastron in the same way that GR effects do, which will affect the distribution of hotspots on secular timescales (for example, see Raymond et al 2014 and references within).

Most studies of exoplanet spin states use tidal evolution models that neglect perturbations from other bodies (e.g. Correia and Laskar 2004, Dobrovolskis 2007), with the exception of simulations that study the Kozai Lidov mechanism, which requires a stellar mass companion orbiting at high inclination to the ecliptic (e.g. Fabrycky and Tremaine 2007). The evolution of spin and orbital angular momentum under the influence of multiple planets is a significant endeavour, and as such we note it as an important avenue both for future studies of planetary dynamics and their astrobiological implications.

**4.5.3 Radiative Transfer**

This work has been concerned solely with the insolation patterns experienced by the ELP in the 3:2 resonance, and has not considered the consequences for the ELP's atmospheric dynamics. To understand this fully requires the use of Global Circulation Models (GCMs) to simulate the radiative transfer of sunlight through the atmosphere to the planet's surface, the hydrological cycle and the production of clouds, and the resulting hydrodynamics of the atmosphere. This has been studied in varying levels of detail by several groups for the 1:1 resonance (Joshi et al., 1997, Joshi 2003, Edson et al 2011, 2012, Kite et al. 2011, Yang et al., 2013), and it is clear that the atmosphere plays an important role in advecting both heat and water vapour, modifying the outgoing longwave radiation (OLR), and consequently the mean surface temperature distribution. Future work would require us to investigate these effects further.

**4.5.4 Obliquity**

We have assumed throughout that the obliquity of the planet is fixed at zero, as tidal interactions that produce the spin-orbit resonance can also erode any obliquity (as is the case with Mercury). However, this is not generally the case - tidal evolution traps planets in one of a number of Cassini states (Dobrovolskis 2009 and references within), which include both low and high obliquity states. Dobrovolskis (2013) explored the high obliquity case for spin-orbit resonant planets with eccentricities of 0.2, just above the calculated limit for stellar retrograde motion. Future work could investigate the currently unexplored high eccentricity, high obliquity regime for this resonance.

# 5 Conclusions

Resonances in the solar system are relatively commonplace. Spin-orbit resonance is one such observed resonance where the period of rotation and period of revolution of the planet are related by an integer ratio. We map the flux patterns incident on the surface of a planet in 3:2 spin-orbit resonance around an M star, and discuss the potential for photosynthetic organisms to survive in such an environment.

While the model is simplistic in its lack of incorporation of radiative transfer or cloud cover, we show that planets in the 3:2 spin orbit resonance are likely to host biomes localised to preferred longitudes, rather than preferred latitudes as is the case on Earth. The exact location of these preferred regions is sensitive to the relative phase of the orbital and revolutionary periods, as well as the planet's orbital eccentricity. Individual "days" experience vastly different strengths of flux, and organisms will require the ability to remain dormant for timescales as long as the longest measured on Earth, and be able to synchronise their photosynthetic activity with the non-trivial periodicity of the flux patterns received at a particular longitude.

This being said, the 3:2 spin-orbit resonance does not appear to mean that a planet is prima facie uninhabitable. However, further work is needed in this area, including more realistic atmospheric modelling and climate simulation.

## Acknowledgements

SB's contribution is in partial fulfilment of the requirement for a St Andrews MRes. AM acknowledges the support of a STFC studentship. DF and CC gratefully acknowledge support from STFC grant ST/J001422/1. The University of Edinburgh is a charitable body, registered in Scotland, with registration number SC005336. The University of Dundee is a charitable body registered in Scotland, with registration number SC 15096.